\documentclass[conference]{IEEEtran}
\usepackage{subfigure}
\usepackage{macros}
\usepackage{todonotes}
\usepackage{booktabs}
\usepackage[export]{adjustbox}
\usepackage{nicefrac}
\usepackage{bm}

\makeatletter
\def\ps@IEEEtitlepagestyle{%
  \def\@oddfoot{\mycopyrightnotice}%
  \def\@oddhead{\hbox{}\@IEEEheaderstyle\leftmark\hfil\thepage}\relax
  \def\@evenhead{\@IEEEheaderstyle\thepage\hfil\leftmark\hbox{}}\relax
  \def\@evenfoot{}%
}
\def\mycopyrightnotice{%
  \begin{minipage}{\textwidth}
  \centering \scriptsize
  Copyright~\copyright~2024 IEEE. Personal use of this material is permitted. Permission from IEEE must be obtained for all other uses, in any current or future media, including\\reprinting/republishing this material for advertising or promotional purposes, creating new collective works, for resale or redistribution to servers or lists, or reuse of any copyrighted component of this work in other works by sending a request to pubs-permissions@ieee.org.
  \end{minipage}
}
\makeatother

\IEEEoverridecommandlockouts

\begin{document}

\title{Designing Reliable Virtualized Radio Access Networks}

\author{
   \IEEEauthorblockN{
     Ufuk Usubütün\IEEEauthorrefmark{1},
     André Gomes\IEEEauthorrefmark{2}\IEEEauthorrefmark{3},
     Shankaranarayanan Puzhavakath Narayanan\IEEEauthorrefmark{4},\\
     Matti Hiltunen\IEEEauthorrefmark{4},
     Shivendra Panwar\IEEEauthorrefmark{1}
   }\vspace{0.5em}
    \IEEEauthorblockA{\IEEEauthorrefmark{1}\textit{New York University}, NY \hspace{3mm} \IEEEauthorrefmark{2}\textit{Virginia Tech}, VA \hspace{3mm} \IEEEauthorrefmark{3}\textit{Rowan University}, NJ \hspace{3mm} \IEEEauthorrefmark{4}\textit{AT\&T Labs Research}, NJ \vspace{0.5em}}
    {Emails: \{usubutun, panwar\}@nyu.edu, gomesa@rowan.edu, \{snarayanan, hiltunen\}@research.att.com}
   \thanks{André Gomes was with Virginia Tech and is now with Rowan University.}
   \thanks{This research was supported by the New York State Center for Advanced Technology in Telecommunications and Distributed Systems (CATT), NYU Wireless, the Cisco University Research Program Fund and by the National Science Foundation (NSF) under Grant No. CNS-2148309 and OAC-2226408.}
}
\IEEEaftertitletext{\vspace{-1.2em}}
\maketitle




\begin{abstract} 
As virtualization of Radio Access Networks (RAN) gains momentum, understanding the impact of hardware and software disaggregation on resiliency becomes critical to meet the high availability requirements of mobile networks. 
Our paper presents an analytical model, using continuous time Markov chains, to study the impact of virtualization and disaggregation on RAN availability.
Our evaluation, assuming typical parameter value ranges for failure and recovery rates, points to containerized platform reliability as a constraint on vRAN availability. We also find that with active-passive replication, increasing hardware replication factor beyond 2 may not bring any benefits unless failover times are reduced. We also compare the reliability of centralized and distributed virtualized central units.

\end{abstract}


\section{Introduction}

Radio Access Network (RAN) resiliency is becoming increasingly critical as we grow more reliant on mobile networks. To provide dependable networks, Mobile Network Operators (MNOs) routinely use advanced network planning strategies like coverage optimization, spectrum redundancy, and optimized scheduling of planned maintenance activities, along with upgrading their network equipment for resiliency.
Maintaining or improving  availability is critical as MNOs transition to \ac{vRAN}.

\begin{figure}[]
\centering
  \includegraphics[width = \linewidth,keepaspectratio]{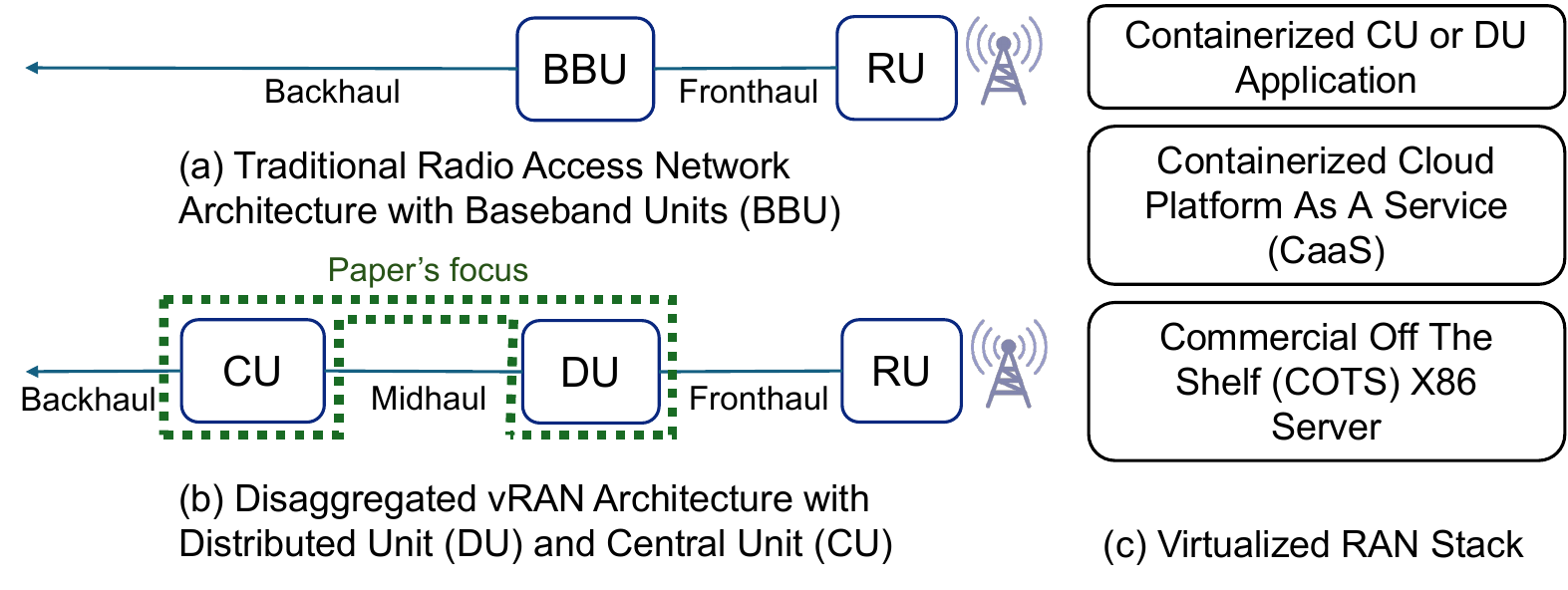} 
\caption{Traditional vs Virtualized Disaggregated Radio Access Network (RAN) Architectures.}
\label{fig:example-ran-architecture}
\vspace{-0.16in}
\end{figure}

The \ac{vRAN} architecture disaggregates the \ac{BBU} into two components: a \ac{DU} that is responsible for processing lower network layers (e.g., PHY, MAC) and a \ac{CU} that is responsible for processing higher network layers (e.g., PDCP, RRC). 
\Fig{example-ran-architecture} contrasts this architectural change from traditional RAN to \ac{vRAN}. 
vRAN also takes the path of virtualization as opposed to traditional \acp{BBU}, which rely on purpose-built hardware running highly customized RAN software. In vRAN, the DU and the CU functions are deployed as software applications, often designed as micro-services, that are deployed on containerized cloud platforms (CaaS). 
These are hosted on \ac{COTS} servers potentially provided by different vendors. 
This allows MNOs to leverage increased flexibility in terms of vendor, network deployment 
and management options. 
However, the fundamental architectural changes brought by \ac{vRAN} require a careful consideration of RAN resiliency. 


Prior works have studied RAN resiliency by focusing on the radio link (e.g., through coverage planning \cite{andrews2011tractable}, spectrum dimensioning \cite{gomes2023assessing}, and user offloading 
\cite{yang2017predictive}) and, more recently, on specific vRAN 
components beyond the radio link  (e.g., improving the PHY layer \cite{msr_slingshot} and stackwide \cite{msrAtlas} DU failovers). 
 
In contrast, our work focuses on examining the architectural change introduced by virtualization and disaggregation, and addresses a critical gap in the study of the reliability of vRANs \cite{ranReliability}. 
We develop analytical models to capture the virtualized and disaggregated vRAN architecture and identify reliability bottlenecks.
We use Markov chains to represent the availability-state of each RAN element, including the COTS server, CaaS platform (which we consider jointly with the operating system (OS)) and CU/DU applications. We use time-to-fail and repair statistics, replication factors and failover mechanisms 
applicable to the respective network elements to model the state transitions. Our models focus only on the functions of the traditional BBU that are disaggregated
into CU and DU (highlighted in Fig.~\ref{fig:example-ran-architecture}(b)). We omit the analysis of midhaul 
as it is often realized through the existing standard transport infrastructure supporting the backhaul.
We model both the active-active and active-passive replication and failover modes, and discuss their impact on availability. We also analyze the impact on overall availability when the CU is centralized, serving multiple cell-sites.

To the best of our knowledge, this is the first paper to develop availability models capturing the key elements of vRAN. Our evaluation of the analytical model shows that: (i) OS/CaaS platform reliability is likely to be a key bottleneck, and improving it is critical to achieving higher vRAN availability, (ii) in active-passive replicated systems, increasing hardware replication factor beyond $2$ provides only limited benefits as failover times tend to limit availability and, (iii) centralized CU deployments have the potential to provide high availability as long as the CU availability is carefully calibrated to reduce single points of failure.

\section{Background}\label{sec:background}

This section outlines the types of failures that may affect \ac{vRAN}s and potential ways of mitigating their impact. 



\subsection{Nature of Failures and Recovery in vRAN}\label{sec:back-nature}

\begin{figure}[]
\centering
  \includegraphics[width = \linewidth,keepaspectratio]{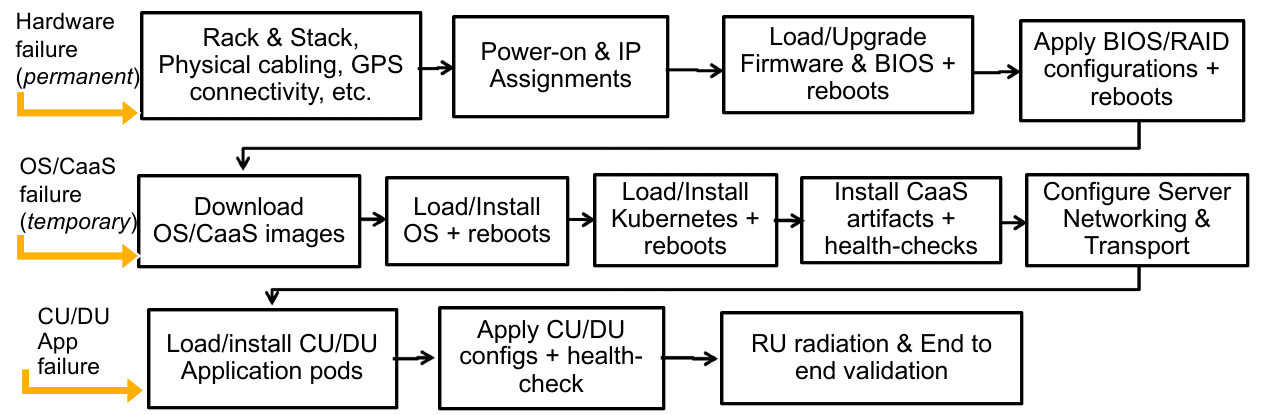} 
\caption{Recovering from failures in a vRAN.}
\label{fig:repalcement}
\vspace{-0.13in}
\end{figure}
The virtualized nature of vRAN requires different recovery actions depending on the nature of the failure. Figure~\ref{fig:repalcement} shows the steps involved in replacing a failed server at the cell-sites -- from physically connecting the server to end-to-end validation that checks if the RAN is ready to serve traffic. Note that some recovery actions may skip certain steps, and each step can take from a few seconds (e.g., applying configurations) up to tens of minutes (e.g., loading BIOS/firmware and multiple reboots). 

For simplicity, we group the failures into three categories: (i) hardware (that we term \textit{permanent}), (ii) OS/CaaS software (that we term \textit{temporary}), and (iii) CU/DU application failures. The duration of hardware failure outages are dominated by a team having to drive to the location of failure and are usually resolved within roughly $10$ hours. For non-permanent failures, recovery time is dominated by the steps outlined in Fig.~\ref{fig:repalcement}. We expect its duration to be of the order of tens of minutes for OS/CaaS (\textit{temporary}) failures and of the order of minutes for CU/DU application failures. However, as data on commercial implementations of vRAN is not available to date to provide precise numbers, we consider wide ranges of values when we evaluate the dynamics of recovering from OS/CaaS software or CU/DU application failures, with specific ranges presented in \Sec{evaluation}.




\subsection{Replication of RAN Elements for Resiliency}\label{sec:back_types}
\begin{figure}[]
\centering
  \includegraphics[width = \linewidth,keepaspectratio]{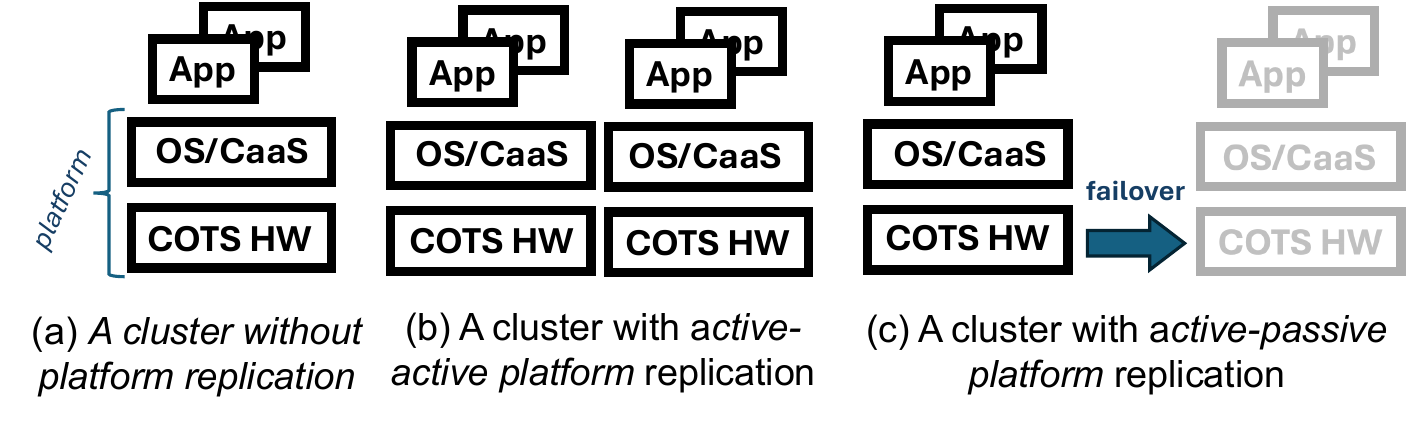} 
  \vspace{-0.23in}
\caption{Different modes of platform replication. Note that in our models we assume that OS/CaaS software is not independently replicated.}
\vspace{-0.1in}
\label{fig:modes}
\end{figure}

vRAN decouples CU/DU applications and the hardware platform, enabling them to be replicated for both resiliency and scaling their capacity (e.g., Fig. \ref{fig:modes}). This allows both the CU and DU to have one or more application instances running on one or more servers. While many combinations of replication are possible, we focus on two commonly adopted replication models: (i) one or more servers with a single CU/DU application instance on each server and (ii) more than one application instance on each server, reflecting the scenario where the CU/DU application instances serve different carrier configurations (e.g., mid-band vs. low-band). 

Finally, the replicas of the \textit{platform} (defined in the next section along with \textit{cluster}) can be either in:


\noindent \emph{\textbf{Active-Active mode:}} Here the incoming traffic to a network function can be processed by any of the replicas independently; therefore the function would be assumed unavailable only when all of its replicas are unavailable. 


\noindent \emph{\textbf{Active-Passive mode:}} Here the incoming traffic coming to a network function would be processed only by a single active replica, with the other passive replicas remaining in a stand-by or idle state. A fail-over mechanism detects the failure of the active replica and elects one of the passive replicas to take over as the active replica \cite{msrAtlas}. Therefore, the failure of an active replica will render the network function unavailable until 
the failover mechanism transitions the function to a previously passive replica. 
The failure of a passive replica will not lead to an outage. 

\section{Modeling vRAN Availability} \label{sec:modeling}

{\color{blue}





}

In this section we first provide necessary definitions and assumptions for modeling. We then model the availability of a DU/CU. We finally consider the reliability implications of a centralized vs. distributed CU.

\subsection{Definitions}

We quantify network resiliency through \textit{\textbf{availability}}, i.e.,
the fraction of time a network is ready to serve incoming network traffic.
 We characterize failure dynamics using the standard metrics: \textbf{\textit{\ac{MTTF}}}, \textbf{\textit{\ac{MTTR}}} \cite{reliabilityText}, 
and a metric called \textbf{\textit{\ac{MTFO}}} that we define as the average time it takes for a system to failover from a failed replica to a functioning one. In this work, we assume that the time-to-failure, time-to-recover and time-to-failover are exponentially distributed. This assumption allows us to build our models using continuous time Markov chains (CTMC) inspired by classic machine-repairman problems 
\cite{machineRepair}. 

We define a \textbf{\textit{platform}} to refer to the combination of a COTS server and an OS/CaaS environment. We then define a \textbf{\textit{cluster}} to contain all replicas of a platform with all CU/DU application replicas running on top of them. Fig. \ref{fig:modes} demonstrates three examples of clusters without and with different modes of platform replication. Therefore, the CU and DU shown in Fig. \ref{fig:example-ran-architecture}(b) represent by one distinct cluster each in our terminology. 

We initially set up models to calculate the availability of platforms and CU/DU applications separately. We then combine the two models to obtain the availability of a cluster. We finally evaluate the availability of the vRAN by taking into account the geographical placement of the CU and DU and obtain a metric representing the number of unavailable cell-sites. 



\subsection{Modeling CU/DU Application Availability}

\begin{figure}[]
    \centering
    \subfigure[][Single]{ \includegraphics[height=0.13\linewidth]{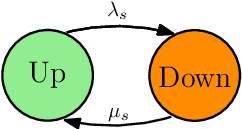} \label{fig:single_sw}}
    \subfigure[][Replicated]{ \includegraphics[ height=0.15\linewidth ]{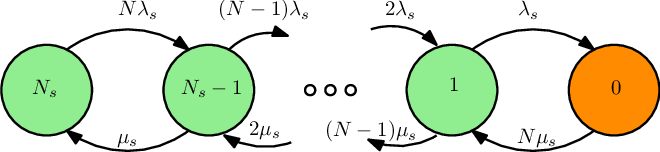} \label{fig:multi_sw}}
    \vspace{-0.09in}
    \caption{CU/DU application availability}
    \vspace{-0.13in}v
\end{figure}

Let the MTTF and MTTR for CU/DU application be MTTF$_s=1/\lambda_{s}$ and  MTTR$_s=1/\mu_{s}$. Note that, for convenience, all notation used in this paper is listed in \Tab{notation}. We model a single, non-replicated, CU/DU application using the two state CTMC in Fig. \ref{fig:single_sw}. 
 The stationary probabilities of the two states are as follows \cite{machineRepair}:
\begin{align}
    \Pr(\text{Up})=\frac{\mu_{s}}{\lambda_{s} + \mu_{s}}, \quad \Pr(\text{Down})=\frac{\lambda_{s}}{\lambda_{s} + \mu_{s}}.
\end{align}

We assume that vRAN applications are built to be cloud-native and are deployed in virtualized environments where a controller is in charge of detecting failures and restarting software instances.
This environment allows us to assume that all replicas of applications fail and recover independently. We therefore formulate replicated applications as $N_s$ independent systems, where $N_s$ is the replication factor, as shown in Figure 4(b), where the states represent the number of up replicas. This yields a binomial distribution $p_i$ of having $i \in \{0,1,\hdots,N_s\}$ replicas up, $p_i = {N_s \choose i} (\frac{\mu_{s}}{\lambda_{s} + \mu_{s}})^{i} (\frac{\lambda_{s}}{\lambda_{s} + \mu_{s}})^{N_s - i} $
and the system stays available as long as there is at least one available replica\cite{tasi}. Therefore, the availability of a CU/DU application with replication factor $N_s$ is equal to $f_s(N_s) = 1-p_0$.

\subsection{Modeling Platform Availability}\label{sec:model:plat}

\begin{figure}[]
    \centering
    \includegraphics[width=0.65\linewidth]{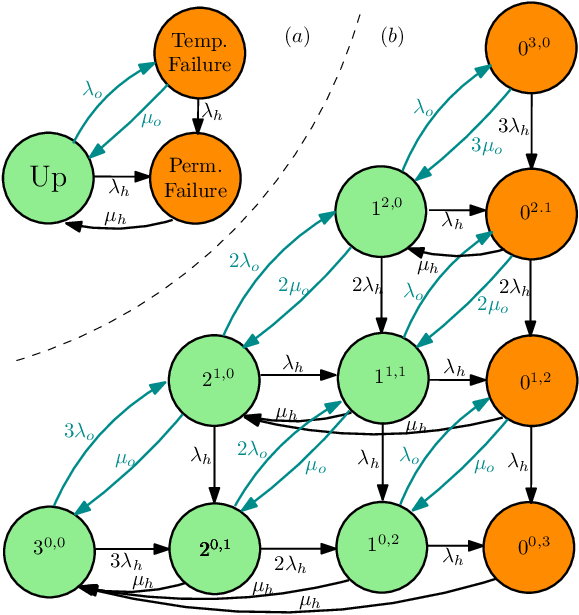} \label{fig:aa_model}  
    \caption{Platform with (a) no replication and (b) \textit{active-active} replication, $N_h=3$.}
    \label{fig:diag_aa}
    \vspace{-0.13in}
\end{figure}

As introduced in \S \ref{sec:back-nature}, the platform can suffer both from temporary and permanent failures. As the recovery dynamics of the two types of outages are related, we model them together. Let the MTTF and MTTR of temporary failures be MTTF$_o = 1/\lambda_o$ and MTTR$_o =1/\mu_o$ and that of permanent failures be MTTF$_h=1/\lambda_h$ and MTTR$_h=1/\mu_h$, respectively. Fig. \ref{fig:diag_aa}(a) illustrates the dynamics of the non-replicated platform: temporary and permanent failures occur and are repaired at their own respective rates. Note that the model also shows the case when a permanent failure occurs during a temporary failure with the downward transition. 

To represent cases with replication, we provide different models for active-active and active-passive modes:

\noindent\textit{\textbf{Active-active replication of the platform:}} 
As any replica can be involved in both temporary and permanent failures with different recovery dynamics, states capturing all combinations of occurences of such outages are needed. The set 
$\mathcal{S}_A=\{a^b : \forall a,b \in \mathbb{Z^+}, \,  a+b \leq N_h \}$
contains all states for a replication factor of $N_h$ and where $\mathbb{Z^+}$ stands for non-negative integers.  
In the figures, a third redundant index, $c$, is introduced to make it easier to interpret the states $a^{b,c}$, as explained next. 
Index $a$ represents the number of functional replicas, $b$ is the number of replicas with temporary failures and $c$ is the the number of those with permanent failures. The third index $c$ can be found by solving for $a+b+c = N_h$. In Figure 5(b),  
the rates $r(.,.)$ between pairs of state are as follows:
\begin{align}
    r(a^{b},(a-1)^{b+1}) &= a \lambda_o \label{eq:os_fail}\\
    r(a^{b},(a+1)^{b-1}) &= b \mu_o \label{eq:os_repair} \\
    r(a^{b},(a-1)^{b}) &= a \lambda_h \label{eq:hw_fail}\\
    r(a^{b},a^{b-1}) &= b \lambda_h \label{eq:hw_fail_s}\\
    r(a^{b},(N_h-b)^{b}) &= \mu_h  \quad , \text{ if } a + b < N_h.\label{eq:hw_repair}
\end{align}
Observe that the rates of failures are scaled by the number of replicas that are functional (index $a$), which can experience failure (Eq. \ref{eq:os_fail},\ref{eq:hw_fail}), and that the rates of recovery from temporary failures are likewise scaled by the number of replicas that are experiencing a temporary outage (index $b$) (Eq. \ref{eq:os_repair}). Likewise, downward permanent failure transitions are also scaled by the number of replicas in temporary outage (Eq. \ref{eq:hw_fail_s}). 
Finally, we assumed that the repair times for permanent, i.e., hardware, failures are dominated by the time it takes for a service team to be dispatched to the location of the failure. Therefore, all such repair transitions lead to the respective state where no permanent failures remain (Eq. \ref{eq:hw_repair}).


We denote the steady state probabilities $\pi(s)$ for each state $s \in \mathcal{S}_A$. For a given state space and its infinitesimal generator matrix $\textbf{Q}$, one can find the embedded discrete time Markov chain (DTMC), solve the stationary state probabilities of the DTMC numerically and scale the solution back such that it provides the steady state solution to the original CTMC \cite{markovBookVol2}. Using this standard procedure, we populate $\textbf{Q}$ 
and solve for the stationary probabilities $\pi(s)$. 

The probability of outage $p_{a}$ can then be found by summing over the steady state probabilities of outage states: $\sum^{N_h}_{b=0} \pi(0^{b}) = p_{a}$. Therefore, the availability of an active-active replicated platform with replication factor $N_h$ is equal to $f_{p,a}(N_h) = 1-p_{a}$.

\begin{figure}
\centering
    \includegraphics[width=0.7\linewidth]{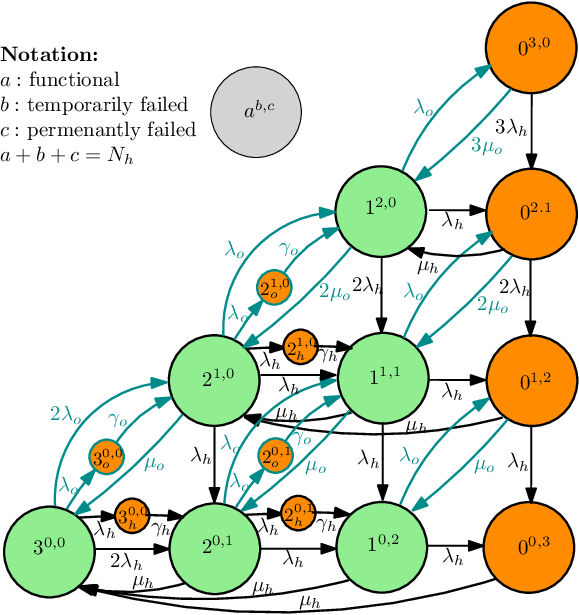} \label{fig:asb_model}
    \caption{\textit{Active-passive} replicated platform, $N_h=3$.}
    \label{fig:diag_asb}
    \vspace{-0.13in}
\end{figure}

\noindent\textit{\textbf{Active-passive replication of the platform:}} To model failovers in the active-passive case, let the MTFO for temporary and permanent failures be MTFO$_o=1/\gamma_o$ and MTFO$_h=1/\gamma_h$ respectively. We introduce new failover states $a_o^{b}$ and $a_h^{b}$ in order to represent the new states during failovers. Formally, the set 
$\mathcal{S}_F = \{a_k^b : \forall k \in \{o,h\}, \, \forall a,b \in \mathbb{Z^+}, \, 2 \leq a ,\, a+b \leq N_h \}$
contains all new states. 
Therefore $\mathcal{S}_P= \mathcal{S}_A \cup \mathcal{S}_F$ contains all states for the active-passive model.
Fig. \ref{fig:diag_asb} depicts the model with the same redundant index $c$.
The model accommodates one major change: Only one of the functioning replicas is actively serving at a time, while others are in standby mode. Therefore, a failure at an active replica triggers a failover that leads to temporary outage. However, a failure in a passive, yet functional, replica does not cause an outage but only reduces the number of available functional components. Finally, there can be no failover if there are no functional passive replicas left. State transition rate equations describing the model are as follows:
\begin{align}
    r(a^{b},(a-1)^{b+1}) &= \max\{a-1,1\} \lambda_o \label{eq:fail_os_p1}\\
    r(a^{b},a_o^{b}) &= \lambda_o \, , \quad r(a_o^{b},(a-1)^{b}) = \gamma_o \label{eq:fail_os_p2}\\
    r(a^{b},(a-1)^{b}) &= \max\{a-1,1\} \lambda_h \label{eq:fail_hw_p1}\\
    r(a^{b},a_h^{b}) &= \lambda_h \, , \quad r(a_h^{b},(a-1)^{b}) = \gamma_h \label{eq:fail_hw_p2}
\end{align}
Equations \ref{eq:os_repair}, \ref{eq:hw_fail} and \ref{eq:hw_repair} remain unchanged from the active-active case and also describe this model. In line with the changes discussed, equations \ref{eq:fail_os_p1} and \ref{eq:fail_hw_p1}  capture the failures incurred by passive replicas, whereas equations \ref{eq:fail_os_p2} and \ref{eq:fail_hw_p2} capture failures incurred by the serving replica and its failover procedure.

The stationary probabilities $\pi(s)$ can be found using the the same methodology as described for the active-active case for all $s \in \mathcal{S}_P$. This time, the outage will be the sum of states with no surviving replicas and the failover states, i.e., $p_{p} = \sum^{N_h}_{b=0} \pi(0^{b}) + \sum_{s \in \mathcal{S}_F} \pi(s)$ and the availability will be $f_{p,p}(N_h) = 1 - p_{p}$.

\subsection{Modeling Cluster Availability}\label{sec:cluster}




We obtain cluster availabilities $f_{c,a}$ and $f_{c,p}$ for a cluster with active-active or active-passive platform replication as:
\begin{align}
    f_{c,a} &= f_{p,a}(N_h) \cdot f_s(N_s \cdot N_h) \label{eq:cluster_av_aa}\\
    f_{c,p} &= f_{p,p}(N_h) \cdot f_s(N_s) \label{eq:cluster_av_ap}
\end{align}

In case of active-active platform replication, we assume application instances across different platforms can coordinate, therefore the number of application replicas are scaled by the number of platform replicas. This is visually illustrated in Fig. \ref{fig:modes}. 
The availability of the CU ($f_{CU}$) and DU ($f_{DU}$) can now be calculated using one of the two options for cluster availability above in equations \ref{eq:cluster_av_aa} and \ref{eq:cluster_av_ap}.


\subsection{Modeling vRAN Availability}\label{sec:model:horiz}

\begin{figure}[]
\centering
  \includegraphics[width = 0.99\linewidth,keepaspectratio]{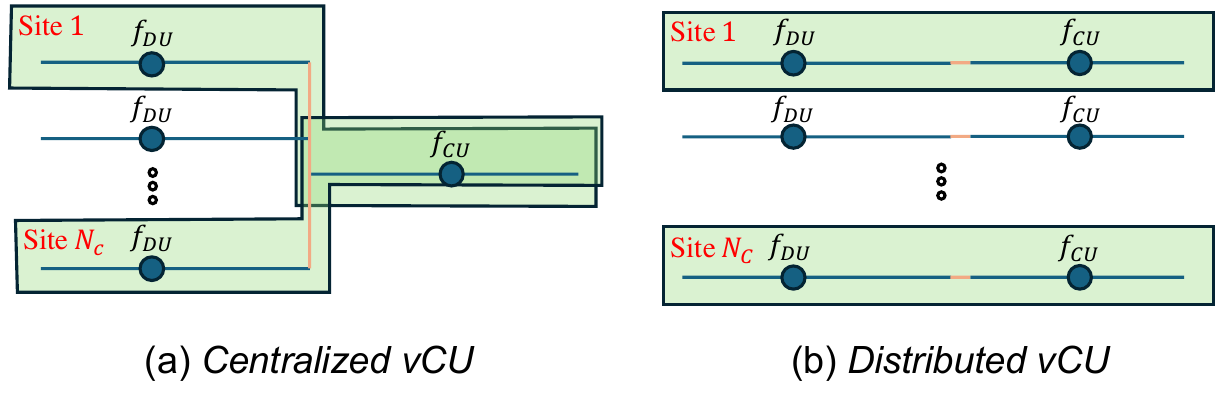} 
\vspace{-0.2in}
\caption{Cell Sites Dependencies with respect to CU placement.} 
\label{fig:CvD}
\vspace{-0.13in}
\end{figure}

The disaggregated vRAN architecture allows the CU and DU to be either co-located or be placed at different physical locations. Here, we focus on two cases of centralized and distributed CUs. Having a centralized CU indicates that multiple DUs at different locations would be served by that single centralized CU, as shown in Fig. \ref{fig:CvD}. 
Let us compare two deployments with $N_c$ cell sites with and without a centralized CU. We assume independence between all instances of DUs and CUs. Let $X_c$ and $X_d$ be the discrete random variables representing the number of unavailable cell sites for centralized and distributed scenarios respectively. 
\begin{align}
    \Pr(X_c=k) = 
    \begin{cases}
        f_{CU}\binom{N_c}{k}(\overline{f_{DU}})^k (f_{DU})^{N_c-k} \,,\text{else}\\
        \overline{f_{CU}} + f_{CU} (\overline{f_{DU}})^{N_c} \quad,\text{if}~k = N_c
    \end{cases}
    \end{align}
    \begin{align}
    \Pr(X_d=k) = \binom{N_c}{k}\left(1-f_{DU}f_{CU}\right)^k \left( f_{DU}f_{CU} \right)^{N_c-k}
\end{align}
Interestingly, even though the two distributions above are very different, the expected number of unavailable cells is the same for both, as derived in Appendix \ref{app:derivation}: 
    $\mathbb{E}[X_d] = \mathbb{E}[X_c] =N_c(1-f_{DU}f_{CU})$.
This can be intuitively explained as follows. The probability of no outage for the centralized case $\Pr(X_c=0)$ is higher than for the distributed case $\Pr(X_d=0)$.
However, when the centralized CU does fail, it brings down all the cells it is connected to. 
This is unlike the failure of a distributed CU deployment for which the failure of all $N_c$ cells is extremely unlikely. 
Note that the outage probability of a cell is also identical in both cases and is equal to $1-f_{DU}f_{CU}$. 



\begin{table*}[]
\caption{Glossary: Summary of terms, definitions, and dependencies}
    \vspace{-0.1in}
    \centering
    \begin{tabular}{ccc}
        \hline 
         \textbf{Term} & \textbf{Definition} & \textbf{Dependence}\\
         \hline
         $1/\lambda_s, 1/\mu_s$   &  CU/DU Application Mean-Time-to-Fail / Mean-Time-to-Recover  & - \\ 
         $ 1/\lambda_o, 1/\mu_o$ & OS/CaaS Software Mean-Time-to-Fail / Mean-Time-to-Recover  &  - \\
         $1/\lambda_h, 1/\mu_h$  & Hardware Mean-Time-to-Fail / Mean-Time-to-Recover  &  - \\
         $1/\gamma_h$  & Hardware Mean-Time-to-Failover  &  - \\
         $1/\gamma_o$  & OS/CaaS Software Mean-Time-to-Failover  &  - \\
         $N_s, N_h$  & Replication Factor for CU/DU Application / Hardware  &  - \\
         $f_s$  & CU/DU Application Availability  &  $\lambda_s, \mu_s, N_s$ \\
         $f_{p,a}$  & Platform Availability with \textit{Active-Active} Replication  &  $\lambda_o, \mu_o, \lambda_h, \mu_h, N_h$ \\
         $f_{p,p}$  & Platform Availability with \textit{Active-Passive} Replication  &  $\lambda_o, \mu_o, \lambda_h, \mu_h, N_h, \gamma_o, \gamma_h$ \\
         $f_{c,a}$  & Cluster Availability with \textit{Active-Active} Replicated Platform  & $f_{p,a}$, $f_s$\\
         $f_{c,p}$  & Cluster Availability with \textit{Active-Passive} Replicated Platform  & $f_{p,p}$, $f_s$\\
         $f_{CU}, f_{DU}$  &  Availability of CU / DU  & $f_{c,a}$ or $f_{c,p}$\\
         $N_c$  & Number of Cell Sites  &  - \\
         $X_c, X_d$ & Number of Unavailable Cell Sites in Centralized / Distributed CU Settings & $f_{DU}$, $f_{CU}, N_c$\\
         \hline 
    \end{tabular}
    \label{tab:notation}
    \vspace{-0.2em}
\end{table*}
\section{Evaluation and Results}
\label{sec:evaluation}

This section reports on vRAN availability using the models introduced in \Sec{modeling}. For the evaluation, we consider the values in \Tab{pub_values} to assess platform, cluster, and vRAN availability. Given the current absence of studies reporting on the failure and recovery rates of vRAN implementations, we refer to typical value ranges from related domains (see Table II for references) and evaluate the impact of these parameter values on the vRAN availability. We evaluate our models with a range of parameter values to reflect the variable duration of failure and recovery times discussed in \Sec{back-nature}.  

\begin{table}[]
\caption{Evaluation parameter value ranges} 
    \centering
    \footnotesize
    \begin{tabular}{ccc}
        \hline 
         \textbf{Parameter} & \textbf{Value} & \textbf{Reference}\\
         \hline
         MTTF$_h$   &  12 - 33 \si{years} (3\% - 8\%)*  & \cite{server-failure1, server-failure0, 8023108} \\ 
         MTTR$_h$ & 10 \si{hours}  & \S \ref{sec:back-nature}\\ 
         MTTF$_o$ & 17 - 70 \si{days}  & \cite[Table 10]{matias2014empirical}\\ 
         MTTR$_o$ &  0.5 \si{\min} - 1.5 \si{hours}  & \cite{os-startup}, \S \ref{sec:back-nature} \\ 
         MTFO & 0.5 - 10 \si{\min} & \S \ref{sec:back-nature} \\ 
         MTTF$_s$ & 7 - 52 \si{days} & \cite{software-failure} \\ 
         MTTR$_s$ & 1s - 30 \si{\min}  & \cite{container-restart-time, kubernetes-liveness-probe, startup-time-containers}, \S \ref{sec:back-nature} \\ 
         \hline 
    \end{tabular}
    \begin{tabular}{l}
    * shows the equivalent annualized failure rate  for hardware. The MTFO$_o$\\
    and MTFO$_h$ values used are the same and correspond to the value shown\\
    for MTFO in this table.
    \end{tabular}
    \label{tab:pub_values}
    \vspace{-0.13in}
\end{table}

\subsection{Platform Availability}  

\begin{figure*}[]
\centering
  \subfigure[b][]{ \includegraphics[width=0.32\linewidth,keepaspectratio]{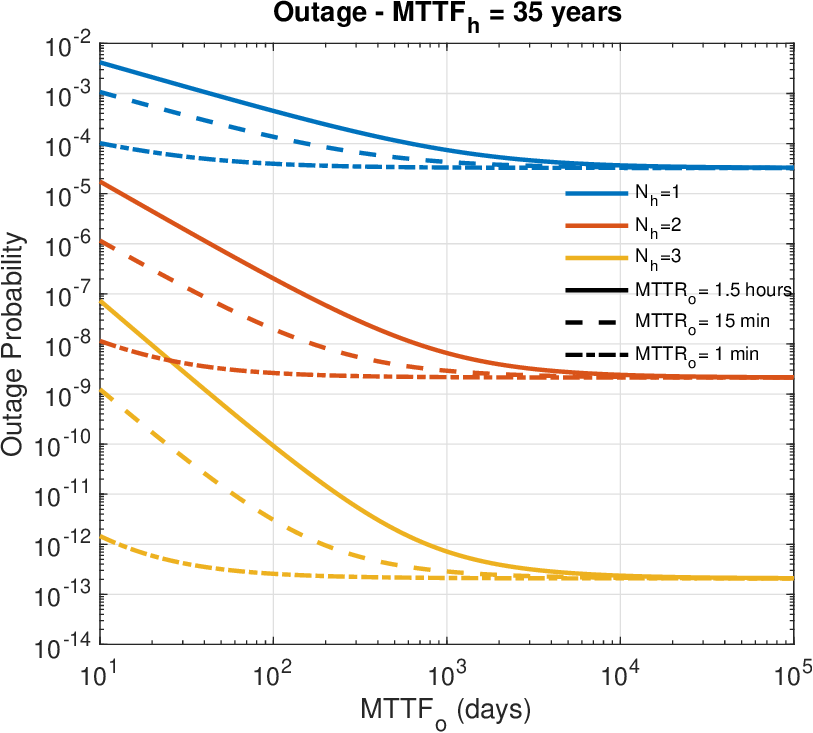}\label{fig:aa2}}
  \subfigure[b][]{ \includegraphics[width=0.32\linewidth,keepaspectratio]{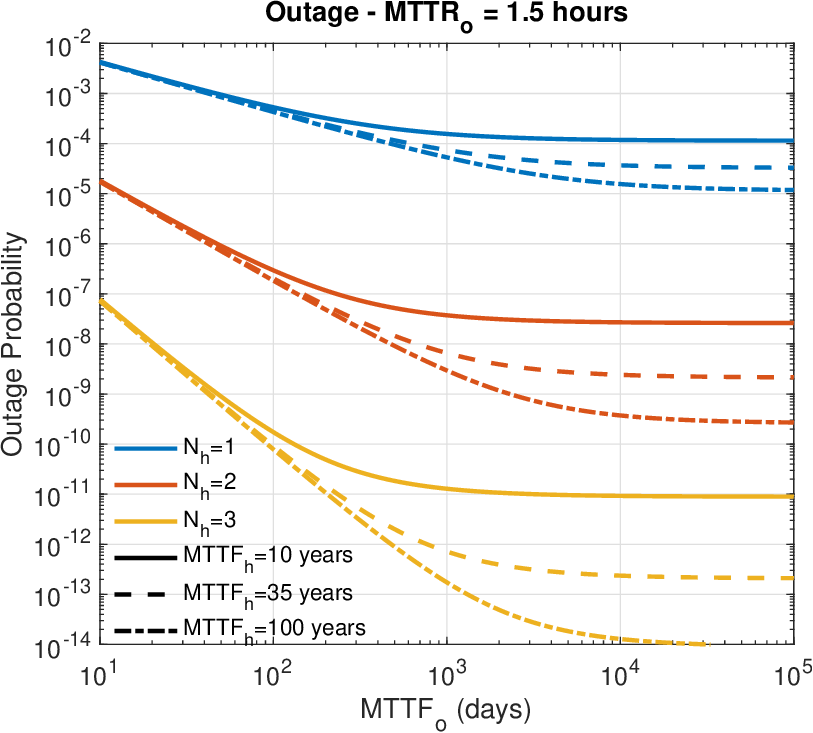} \label{fig:aa1}}
  \subfigure[b][]{ \includegraphics[width=0.32\linewidth,keepaspectratio]{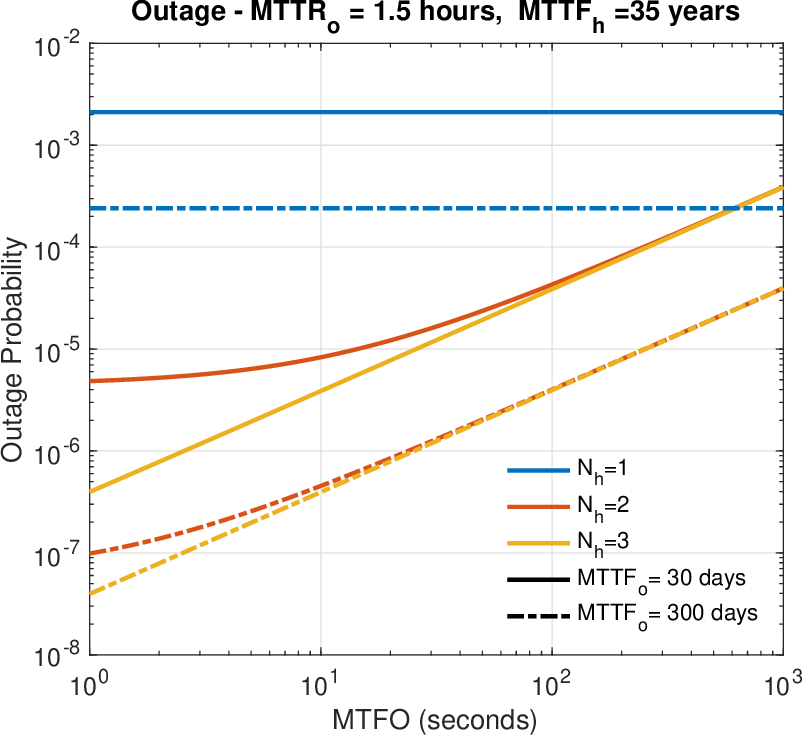}\label{fig:asb}}
  \vspace{-0.13in}
\caption{Platform availability in (a) active-active replication with MTTF$_h$ = 35 \si{years}, (b) active-active replication with MTTR$_o$ = 1.5 \si{\hour},  and (c) active-passive replication with MTTR$_o$ = 1.5 \si{\hour} and MTTF$_h$ = 35 \si{years}. MTTR$_h$ = 10 \si{hours} for all scenarios.}
\label{fig:eval}
\vspace{-0.13em}
\end{figure*}

We start with the platform availability (as modeled in \S \ref{sec:model:plat}) in an active-active replication scenario.  
In \Fig{aa2}, we explore the effect of different combinations of MTTR$_o$ and MTTF$_o$ for fixed hardware failure and repair rates to investigate the impact of OS/CaaS outages on the platform availability. 
Observe that platform availability does not improve further after \ac{MTTF}$_o > 10^3$ \si{days}. 
However, in practice, OS/CaaS can have lower MTTF$_o$ (e.g., \cite{matias2014empirical}), which implies that platform availability can be significantly impacted by failures in the OS/CaaS.
In such cases, timely recovery (\ac{MTTR}$_o \in \{1, 15\}$ \si{min}) can significantly mitigate the impact of failures, reducing the observed platform outage probability. 

\Fig{aa1} considers the unfavorable scenario of \ac{MTTR}$_o~=~1.5$ \si{hours} to represent longer repair times that may involve reinstalling OS/CaaS environments. 
Interestingly, for \ac{MTTF}$_o \le 10^2$ \si{days}, \Fig{aa1} shows little benefit from using more reliable hardware (i.e., higher \ac{MTTF}$_h$) as OS/CaaS failures dominate the overall outage probability. 
Hardware replication, which also replicates OS/CaaS as each server has its own OS/CaaS environment, can compensate for less reliable OS/CaaS environments at the expense of additional hardware and potentially more OS/CaaS software licenses. 
These results highlight the relevance of carefully selecting reliable OS/CaaS environments to avoid frequent failures that can limit \ac{vRAN} availability.

In \Fig{asb}, we consider a counterpart scenario with active-passive replication to study the impact of failover mechanisms on the platform availability. Notice that there is no failover if $N_h = 1$ (i.e., no replication). 
Interestingly, we observe that the outage probability is insensitive to additional replication beyond $N_h = 2$ for the region \ac{MTFO} $> 10$ \si{\sec}. 
This suggests that investments in fast failover (such as the mechanisms proposed in \cite{msr_slingshot, msrAtlas}) can significantly improve availability. It also shows that when failover is slow, investment in additional hardware replication does not lead to increased availability.

\subsection{Cluster Availability} \label{sec:hw-sw-availability} 

\Tab{aa} reports on the entire cluster availability (i.e., platform plus CU/DU applications, calculated as in \Sec{cluster}) for an active-active replication setting. 
For ease of visualization, we focus on the parameter combinations that lead to six-9s availability, where six-9s means availability in the range $[0.999999, 0.9999999)$. We present the overall availability (\#9s) as well as the corresponding platform (\#9s$_p$) and OS/CaaS (\#9s$_s$) availabilities. 
For the range of parameter values in consideration, the target availability of six-9s can only be achieved by hardware replication (i.e., $N_h \ge 2$). After that, software becomes the availability bottleneck, requiring more reliable OS/CaaS environments (e.g., \ac{MTTF}$_o = 10$~\si{months} needed to achieve seven-9s) and faster \acs{CU}/\acs{DU} recovery mechanisms (e.g., \ac{MTTR}$_s = 5$~\si{\min} needed to meet seven-9s). 

Similarly, \Tab{asb} shows the cluster availability for an active-passive scenario. We consider \ac{MTFO} values of \{10, 100\} \si{\sec}. Notice, however, that six-9s availability is only achievable with an \ac{MTFO} of 10 \si{\sec} even in settings with more reliable hardware, better OS/CaaS recovery times, or replication (see \Tab{asb}, rows 3-4). This highlights, again, the importance of investing in fast failover mechanisms to achieve high availability in deployments with active-passive replication. 
We also see that more CU/DU application replicas ($N_s$) are needed on each platform in this case, as replicas hosted on passive platform replicas are not immediately available. Note that our model does not capture failovers that could be triggered after all CU/DU application replicas on a platform fail. Therefore, for the case when the MTFO is shorter than the MTTR$_s$, our model might provide an underestimate of application availability.



\begin{table}[]
\caption{Probability of Cell Outage and Probability of All Cells Unavailable as CU Outage Probability is Decreased} \label{tab:centCu}
    \vspace{-0.7em}
    \centering
\begin{tabular}{cccc}
\hline
    DU Outage & CU Outage & Cell Outage & All Cells Unavailable\\
   $1-f_{DU}$ &  $1-f_{CU}$ &   $1-f_{DU}f_{CU}$ & $\Pr(X_c=N_c)$  \\ 
   \hline
   $10^{-5}$ & $10^{-5}$ &  $\sim 1.99\cdot 10^{-5}$  & $\sim 10^{-5}$\\
   $10^{-5}$ & $10^{-6}$ &  $\sim 1.10\cdot 10^{-5}$  & $\sim 10^{-6}$\\
   $10^{-5}$ & $10^{-7}$ &  $\sim 1.01\cdot 10^{-5}$  & $\sim 10^{-7}$\\
   \hline
\end{tabular}
\vspace{-0.13in}
\end{table}

\subsection{Assessing Network-wide vRAN Availability}



We have found, in \S \ref{sec:model:horiz}, that the expected number of cells in outage for both centralized and distributed CU cases to be the same and equal to $N_c (1-f_{DU}f_{CU})$. At the same time, we also found a significant probability for all cells to simultaneously experience an 
outage for the centralized CU case. To mitigate this, we reason that since it is likely for centralized CUs to be well provisioned to handle the large volume of traffic resulting from a  centralized architecture, it can also be cost-effectively designed for higher reliability as compared to a CU in a distributed architecture that serves only one cell. In \Tab{centCu} we consider three scenarios where we increase the CU availability in each successive row. We provide the cell outage probability, which is the same for both centralized and distributed CU cases, and the probability of all cells  being unavailable for the centralized CU case (for $N_c > 2$).

We see that improvements in the CU availability leads to a corresponding reduction in the probability of all cells being unavailable. 
If the likelihood of this type of outage is an important concern, then the design of a sufficiently reliable centralized CU can mitigate this issue.



\begin{table*}[]
    \centering
    \caption{Cluster availability in active-active redundancy for different combinations of parameters} \label{tab:aa} 
\begin{tabular}{lll|llllll}
\hline
   \#9s &   \#9s$_p$ &   \#9s$_s$ &   $N_h$ & MTTF$_h$                & MTTF$_o$   & MTTR$_o$         &   $N_s$ & MTTR$_s$   \\
\hline

\hline 
     6 &         6 &         6 &       2 & 10 \si{years}              & 10 months  & 90 \si{\min}           &       1 & 30 \si{\min}   \\
     6 &         6 &         6 &       2 & 10 \si{years}, 100 \si{years} & 1 months   & 15 \si{\min}         &       1 & 30 \si{\min}   \\
     6 &         6 &         8 &       2 & 10 \si{years}              & 10 months  & 90 \si{\min}           &       1 & 5 \si{\min}    \\
     6 &         6 &         8 &       2 & 10 \si{years}, 100 \si{years} & 1 months   & 15 \si{\min}         &       1 & 5 \si{\min}    \\
     7 &         7 &         8 &       2 & 10 \si{years}              & 10 months  & 15 \si{\min}         &       1 & 5 \si{\min}    \\
     7 &         7 &         8 &       2 & 100 \si{years}             & 10 months  & 90 \si{\min}           &       1 & 5 \si{\min}    \\
    \hline
\end{tabular}
\begin{tabular}{c}
 For all cases MTTR$_h = 10$ hours, MTTF$_s = 2$ months.\\
All possible combinations of parameters in a given row result in the same number of nines for the cluster, platform and application respectively.
\end{tabular}
    \vspace{-0.1in}
\end{table*}

\begin{table*}[]
    \centering
        \caption{Cluster availability in active-passive redundancy for different combinations of parameters} 
        \label{tab:asb}
\begin{tabular}{lll|lllllll}
\hline
   \#9s &   \#9s$_p$ &   \#9s$_s$ & $N_h$   & MTTF$_h$              & MTTF$_o$   & MTTR$_o$                  & MTFO                &   $N_s$ & MTTR$_s$   \\
\hline
     \textcolor{black}{5} &         \textcolor{black}{5} &         \textcolor{black}{5} & \textcolor{black}{2, 3}    & \textcolor{black}{10 \si{years}, 100 \si{years}} & \textcolor{black}{10 \si{months} }  & \textcolor{black}{1 \si{\min},  15 \si{\min}, 90 \si{\min} } & \textcolor{black}{100 \si{\sec} }             &      \textcolor{black}{ 2} & \textcolor{black}{30 \si{\min}}    \\
     \textcolor{black}{5} &         \textcolor{black}{5} &         \textcolor{black}{7} & \textcolor{black}{2, 3 }   &  \textcolor{black}{10 \si{years}, 100 \si{years}} & \textcolor{black}{ 10 \si{months}  } & \textcolor{black}{1 \si{\min}, 15 \si{\min}, 90 \si{\min}} & \textcolor{black}{100 \si{\sec}   }            &      \textcolor{black}{ 2 } & \textcolor{black}{5 \si{\min}}     \\
     6 &         6 &         7 & 2, 3    & 10 \si{years}, 100 \si{years} & 10 \si{months}   & 1 \si{\min}, 15 \si{\min}, 90 \si{\min} & 10 \si{\sec}                 &       2 & 5 \si{\min}    \\
     6 &         6 &         8 & 2, 3    & 10 \si{years}, 100 \si{years} & 10 \si{months}   & 1 \si{\min}, 15 \si{\min}, 90 \si{\min} & 10 \si{\sec}                 &       3 & 30 \si{\min}    \\
     \hline
\end{tabular}
\begin{tabular}{c}
 For all cases MTTR$_h = 10$ hours, MTTF$_s = 2$ months.\\
All possible combinations of parameters in a given row result in the same number of nines for the cluster, platform and application respectively.
\end{tabular}
    \vspace{-0.1in}
\end{table*}

\section{Conclusions}\label{sec:conc}



Based on our analysis, we came to three main conclusions about vRAN reliability:
\begin{itemize}
    \item Hardware replication is necessary to achieve higher than five-9s availability.
    \item  OS/CaaS reliability constrains the availability of the vRAN. Therefore increasing its availability and reducing repair times will improve overall vRAN availability
    \item For active-passive systems, we determined  that failover times constrain their reliability.
\end{itemize}
Our analysis also exposed an interesting characteristic of centralized CU deployments: they are more likely to have all cell sites unavailable than distributed CUs, even while the two have the same average cell availability. This weakness can be mitigated by making centralized CUs more reliable, an investment that can be justified given that there are far fewer CUs in a centralized architecture as compared to a distributed architecture.



\bibliography{IEEEabrv, bib.bib}
\bibliographystyle{IEEEtran}

\appendices

\section{Derivation of the Expected Number of Unavailable Cells for Distributed and Centralized CUs}
  \label{app:derivation}

\subsection{The Expected Value of $X_d$:}
\noindent By definition of the binomial distribution:
\begin{align}
    \mathbb{E}[X_d] = N_c (1-f_{DU}f_{CU}).
\end{align}

\subsection{The Expected Value of $X_c$:}
\noindent For simplicity define:
\begin{align}
    f_{CU} &= \alpha, \quad \quad \overline{f_{CU}} = 1- f_{CU} = \beta.\\
    f_{DU} &= p, \quad \quad \overline{f_{DU}} = 1- f_{DU} = q.
\end{align}
Substitute $\alpha, \beta, p$ and $q$ into $\Pr(X_c)$:
\begin{align}
    \Pr(X_c=k) = 
    \begin{cases}
        \beta + \alpha q^{N_c},~\text{if}~k = N_c,\\
        \alpha\binom{N_c}{k} q^k p^{N_c-k},~\text{else}.
    \end{cases}
\end{align}
Define random variable $X'$ such that,
\begin{align}
    \Pr(X' = k) &= \binom{N_c}{k}(q)^k (p)^{N_c-k}.
\end{align}
By definition of the binomial random variable
\begin{align}
    \mathbb{E}[X'] &= N_c q \label{eq:binom_eq}.
\end{align}
We can express $\mathbb{E}[X_c]$ in terms of $X'$ as follows:
\begin{align}
    \mathbb{E}[X_c] &= \alpha \big[ \mathbb{E}[X'] - N_c \cdot \Pr(X' = N_c) \big] + N_c \cdot \Pr(X_c = N_c)\\
    & = \alpha \big[ N_c q - N_c q^{N_c} \big] + N_c \cdot (\beta + \alpha q^{N_c} )\\
    & = N_c ( \alpha q + \beta )\\
    & = N_c ( \alpha (1-p) + 1-\alpha )\\
    & = N_c ( 1- p \alpha )\\
    & = N_c (1- f_{DU}f_{CU}) = \mathbb{E}[X_d].
\end{align}

\end{document}